\documentstyle[prl,multicol,aps]{revtex}

\newcommand \bq {\begin{eqnarray}}
\newcommand \be {\begin{equation}}
\newcommand \eq {\end{eqnarray}}
\newcommand \ee {\end{equation}}
\newcommand \pr {\prime}

\renewcommand{\d}{{\rm d}}
\def\dbarrm {{\mathchar'26\mkern-11mu{\rm d}}}                        %
\begin{document} 
\draft
\title{Entropy production, energy dissipation 
and violation of Onsager relations \\ in the steady glassy state.}
\author{A.E. Allahverdyan$^{1,3)}$ and Th.M. Nieuwenhuizen$^{2)}$}
\address{$^{1)}$ Institute for Theoretical Physics, 
$^{2)}$ Department of Physics and Astronomy,\\ 
University of Amsterdam,
Valckenierstraat 65, 1018 XE Amsterdam, The Netherlands 
\\ $^{3)}$Yerevan Physics Institute,
Alikhanian Brothers St. 2, Yerevan 375036, Armenia }
\maketitle
\begin{abstract}
In a glassy system different degrees of freedom have well-separated  
characteristic times, and are described by different temperatures. 
The stationary state 
is essentially non-equilibrium. A generalized statistical 
thermodynamics is constructed and a universal
variational principle is proposed.
Entropy production and energy dissipation occur at a constant
rate; there exists a universal relation between them, valid
to leading order in the small ratio of the characteristic times. 
Energy dissipation (unlike entropy production) is closely connected
to the fluctuations of the slow degree.
Corrections due to a finite ratio  of  the times are obtained. 
Onsager relations in the context of heat transfer are also 
considered. They are 
always broken in glassy states, except close to equilibrium.
\end{abstract}
\pacs{
PACS: 64.70.Pf, 05.70.Ln, 75.10Nr, 75.40Cx, 75.50Lk}
\begin{multicols}{2}
Statistical thermodynamics is a universal and powerful theory
for describing equilibrium states. 
It was generalized to weakly non-equilibrium states, in an 
approach first started by Onsager, and further developed 
extensively, see e.g. \cite{strat}.

It was recognized long time ago that concepts and methods of statistical
 thermodynamics can also be applied to glassy non-equilibrium states
\cite{jones}.
In such systems the relaxation times depend strongly on temperature. 
When cooling at a proper rate (varying from $10^{-2}$ K/s for window
 glass to $10^{5}$ K/s for metallic glasses) 
the equilibrium relaxation time becomes very large near the 
experimentally defined glassy temperature $T_g$. 
The thus reached metastable state is not in equilibrium but,
 nevertheless, can be described by a generalized thermodynamics,
assigning different temperatures (so-called effective or fictive
 temperatures) to processes with well-separated characteristic times
\cite{jones} -\cite{theolongtermo}. 
In spite of much progress in this area many important questions are
still  not fully understood. 
In particular, it concerns dissipative 
effects.
However, as the steady glassy state is non-equilibrium, there exist 
a constant-rate entropy production, an energy dissipation, 
and a transfer of heat.  Although the physical importance
of entropy production was stressed in the fundamental review
~\cite{jones}, its systematic investigation has been 
continued only very recently \cite{theolongtermo}.

We shall consider the steady non-equilibrium glassy state 
of systems in which the subsystems are coupled to baths at
different temperatures.
The glassiness here is solely a consequence of the assumed
large separation of time-scales of the subsystems.
Our purposes are the following. 
{\it (i)} Derive the glassy stationary statistical distribution
and the corresponding thermodynamics.
{\it (ii)} Propose a general variational principle for glassy
thermodynamics. 
{\it (iii)} Investigate the entropy production and energy dissipation 
in the stationary non-equilibrium glassy state.
{\it (iv)} Show the breakdown of the Onsager relations
for heat transfer.

To adstruct our conclusions, let us introduce 
the simplest glassy system. It  consist of a pair of 
coupled stochastic variables $x_1$, $x_2$ with Hamiltonian $H(x_1,x_2)$,
which interact with different thermal baths and have different 
characteristic time-scales. Such an approach pretends to establish 
all important, necessary ingredients of glassy behavior.
A theory of statistical systems interacting with different thermal
baths  was investigated in \cite{hb}.
The essentially new points of our approach are the large separation 
between characteristic times, and arbitrary difference between 
temperatures.
The (over\-damped) Langevin equations for the dynamics read: 
\begin{eqnarray} 
\label{1}
\Gamma _i \dot{x}_i=&&-\partial _i H
 +\eta _{i}(t),\, \langle \eta _{i}(t)\eta
_{j}(t^{\prime })\rangle =2\Gamma _i T_i\delta _{ij}\delta (t-t^{\pr }),
\nonumber\\
i,j=&&1,2
\end{eqnarray}
where $\Gamma _1$, $\Gamma_2$ are the damping constants, 
and $\partial _i=\partial /\partial x_i$. The Einstein
relation between the strength of noise and the damping constant holds 
in Eq. (\ref{1}) because  the thermal
baths themselves are in equilibrium \cite{gardiner}.
It is assumed that the relaxation time toward the total
equilibrium (where $T_2=T_1$) is much larger than all considered times;
thus for our purposes $T_2$ and $T_1$ are constants.

Hereafter we shall assume that $x_2$ is changing much more slowly 
than $x_1$; this condition is ensured by the condition 
$\gamma =\Gamma _1/\Gamma _2\ll 1$.
Let us first indicate how the stationary distribution can be obtained 
to order $\gamma ^0$, which will
give us the basic formulation of the generalized glassy thermodynamics.
Eqs. (\ref{1}) can be investigated by the method
of adiabatic elimination \cite{gardiner} (Born-Oppenheimer method). 
First Eq. (\ref{1}) for $x_1$ is solved keeping the $x_2$
fixed, valid on relatively short time-scales
where only Eq. (\ref{1}) for $x_1$ is relevant. In this case the 
Langevin equation has the equilibrium distribution 
\begin{equation}
\label{3}
P_0(x_1| x_2)=\frac{1}{Z(x_2)}\exp (-\beta _1 H(x_1, x_2)),
\end{equation}
where $Z(x_2)$ is the partition sum for a fixed value of $x_2$.
$x_1$ can be carried out. 
At quasi-equilibrium of the $x_2$-subsystem
this average should be performed using the distribution (\ref{3}).
In this way we get from Eqs. (\ref{1}) 
a related dynamics for the slow variable, in which the two particle
Hamiltonian $H(x_1, x_2)$
is replaced by the effective one-particle Hamiltonian $-T_1\ln Z(x_2)$. 
We thus have the effective equation of motion
\begin{equation}
\label{4}
\Gamma _2\dot{x}_2=\frac{\partial }{\partial x_2}T_1\ln Z(x_2)+\eta_2(t)
\end{equation}
As the noise is due to a bath at temperature $T_2$, see Eq. (\ref{1}), 
the equilibrium distribution of this equation reads
\begin{equation}
\label{5}
P_0(x_2)=\frac{Z^{T_1/T_2}(x_2)}{{\cal Z}}, \, {\cal Z}=\int
 \d x_2 Z^{T_1/T_2}(x_2).
\end{equation}
The joint distribution of $x_1$ and $x_2$ can now be written as
 $P_0(x_1,x_2)=P_0(x_1 |x_2)P_0(x_2)$.
A similar approach is applied in spin-glasses and other disordered systems 
where $n=T_1/T_2$
is considered as``dynamically generated'' replica number
\cite{sherington}.
Keeping this in mind we now consider the general situation.

{\it (i)} If the state of a system is described by a distribution  
$P(x_1,x_2)=P(x_1 |x_2)P(x_2)$ 
then there exist the {\it general definition}
for the mean energy and entropy \cite{strat}: $U=\langle H\rangle$,
 $S=-\langle \ln P\rangle$. This latter Boltzmann-Gibbs-Shannon
formula corresponds to the general statistical definition of entropy,
 relevant also outside of equilibrium \cite{strat}.
The total entropy decomposes as $S=S_1+S_2$, where
\begin{eqnarray}
\label{9}
&& S_1=\int \d x_2\,P(x_2)[- \int \d x_1 P(x_1|x_2)\ln P(x_1| x_2)], 
\nonumber \\  
&& S_2=-\int \d x_2 P(x_2)\ln P(x_2).
\end{eqnarray}
$S_1$ is the entropy of the fast variable $x_1$, averaged over
the quenched slow variable $x_2$,
and $S_2$ is the entropy of the slow variable itself. 
This general result of the statistical thermodynamics can be
again applied in our case when $P(x_1, x_2)=P_0(x_1, x_2)$. From 
Eqs. (\ref{3}), (\ref{5}), (\ref{9}) an important
relation can be derived which {\it generalizes} 
the usual thermodynamical relation
for the free energy $F=-T_2\ln {\cal Z}$:
\begin{equation}
\label{10}
F=U-T_1S_1-T_2S_2
\end{equation}
This agrees with the expression of the free energy for a glassy system
put forward  previously by one of us~\cite{theoehren,theohammer}.
In that approach the equivalent of $T_2$ is the dynamically generated
effective temperature, while here it is the temperature of a bath.
The first and second laws of thermodynamics take the form
\begin{equation}
\label{13}
\d U=\dbarrm Q + \dbarrm W\leq T_1\d S_1+T_2 \d S_2 +\dbarrm W,
\end{equation}
where $\dbarrm W$ is the work which is done on the system by external
 forces, and the equality in Eq. (\ref{13}) is realized for a
 reversible process. 
Eqs. (\ref{10}-\ref{13}) are the manifestation of the glassy 
thermodynamics which generalize the usual one to the case of 
{\it non-equilibrium} systems with well-separated time-scales. 
Here they have been obtained from Langevin equations under the sole
assumption of a separation of time scales.

{\it (ii)} Let us indicate how a variational principle can be obtained 
from a more general consideration. The usual Gibbs
distribution for homogeneous
equilibrium states can be obtained either from maximizing the entropy, 
keeping energy fixed, or from minimizing  the energy, keeping the
entropy fixed. For the glassy state, which is non-homogeneous 
and out of equilibrium, one can minimize
the mean energy, keeping both entropies $S_1$ and $S_2$ fixed
(somewhat similar to the microcanonical approach). 
Following the standard method we should minimize,
with respect to $P(x_2)$ and $P(x_1|x_2)$, the Lagrange function
\begin{eqnarray}
\label{99}
&& {\cal L}= \int \d x_1\d x_2 P(x_1,x_2)H+
T_2\int \d x_2 P(x_2)\ln P(x_2)
\nonumber \\  
&& +T_1\int \d x_2\,P(x_2)\int \d x_1 P(x_1|x_2)\ln P(x_1| x_2),
\end{eqnarray}
where $T_1$ and $T_2$ are Lagrange multipliers, and normalize the
solutions.
We then recover Eqs. (\ref{3}), (\ref{5})
but now on the basis of more general variational principle.
This clearly demonstrates the conceptual
difference
compared to the usual (local-equillibrium) thermodynamics. 

{\it (iii)} Due to a difference  between $T_1$ and $T_2$ there is
constant current of heat through the system. 
This implies a constant production
of entropy and dissipation of energy. We investigate
these effects taking into account possible $\gamma ^2$
corrections. 
The Fokker-Planck equation which corresponds to Eqs. (\ref{1}) reads
\begin{eqnarray}
\label{16}
&&\partial _t P(x_1,x_2;t) + \sum_{i=1}^2 \partial _i J_i =0, 
\nonumber \\
&&\Gamma _i J_i= P(x_1,x_2;t) \partial _i H +T_i
\partial _i P(x_1,x_2;t)
\end{eqnarray}
where $J_1$, $J_2$ are the currents of probability.
The stationary probability distribution can be expressed as 
\begin{equation}
\label{17}
P_1(x_1, x_2)=P_0(x_1, x_2)(1-\gamma A(x_1, x_2))+{\cal O}(\gamma^2),
\end{equation}
The boundary conditions are, as usual, that $P(x_1,x_2)$ and its
 derivatives vanish at infinity. $A$ is obtained
from the stationarity condition $\partial _tP=0$, 
taking into account the orthogonality condition $\int dx_1dx_2AP_0=0$,
and consistency with ${\cal O}(\gamma^2)$ terms.
The general expression for $A$ is rather lengthy, 
but for a concrete model it is given in Eq. (\ref{kk1}).
In the first order of $\gamma $ the steady currents are given by 
\begin{equation}
\label{19}
J_1=\gamma\frac{T_1}{\Gamma _1}P_0\,\partial _1 A, \qquad
J_2=\gamma\frac{T_1-T_2}{T_1\Gamma_1}P_0\, \delta F_2 ,
\end{equation}
Notice that for $J_2$ the object $A$ is not needed, but only
\begin{equation}
\label{19a} 
\delta F_2 =
-\partial _2 H +
\int \d y P_0(y |x_2)\partial _2 H (y,x_2),
\end{equation}
being the difference between the force acting on the second subsystem
and its conventional mean value obtained by averaging over the fast 
degree of freedom. Therefore some further results can be derived without
knowledge of $A$, though it is needed for consistency checks and
$\gamma ^2$-corrections.
The change of total entropy reads
\begin{equation}
\label{20}
\dbarrm S_{tot}=\d S + \dbarrm S_{b,1} +\dbarrm S_{b,2}=\d S -\beta _1 
\dbarrm _1 Q -\beta _2 \dbarrm _2 Q
\end{equation}
where $S=S_1+S_2$ is the entropy of the system defined by (\ref{9}), 
$S_{b,1}$, $S_{b,2}$ are the entropies of the corresponding thermal
baths, and $\dbarrm _1 Q$, $\dbarrm _2Q$ are the amounts of heat
obtained by the system from the thermal baths. Of course, 
from the conservation of energy we have 
$\dbarrm _i Q+\dbarrm Q_{bath,i}=0$, while 
$\dbarrm Q_{b,i}=T_i~\dbarrm S_{b,i}$ holds because the baths
are  in equilibrium. Further, the expression  
\begin{equation}
\label{urn1}
\dot{Q}_i\equiv \frac{\dbarrm _i Q}{dt}=-\int \d x_1\d x_2 H(x_1,x_2) 
\partial _i J_i
\end{equation}
can be obtained from (\ref{16}).
The entropy and the mean energy of the stationary state are constant:
$\dot S=0$, $\dot{Q}_1+\dot{Q}_2=0$.
Nevertheless, there exists a constant-rate transfer of entropy 
to the outside world (the thermal baths),
 and a stationary flux of heat through the system:
\begin{equation}
\label{f1}
\dot{S}_{tot}= (\beta _1-\beta _2)\dot{Q}_2. 
\end{equation} 
 
If the system is in a non-equilibrium steady state then work 
should be done to keep it there.
It is just the work needed for creating macroscopic 
currents inside the system. To illustrate this thesis, we can employ
relation (\ref{13}) for constant $T_1$, $T_2$, divide it by $\d t$
and write it as $\dot{F}= \dot{W} -\dot{\Pi}$.
The positive quantity 
$\dot{\Pi}$ is the energy dissipated per unit of time. 
Using Eq. (\ref{16}) we get 
\begin{eqnarray}
\label{23}
&&\dot{F}= (T_1-T_2)\int \d x_1\d x_2 J_2\partial _2 
\ln P(x_1| x_2) -\nonumber \\
&&\int \d x_1\d x_2 P(x_1,x_2)\sum _{i=1}^2\frac{1}{\Gamma _i}
(\partial _i H +T_i\partial _i \ln P )^2
\end{eqnarray}
The last term in the right-hand side is nothing else but 
the energy dissipated per unit of time; it can be written alternatively as 
the sum of energy dissipation driven by the corresponding thermal baths:
$\dot\Pi=\sum_{i=1}^2(T_i\, \dbarrm_i S- \dbarrm_i Q)/\d t$
(recall that $\dbarrm _i Q$ is the heat obtained from the thermal
 bath $i$, and ~$\dbarrm _i S$ is 
the change of system's entropy induced by this thermal bath).
The first term in the right-hand side of Eq. (\ref{23}) should be 
associated with the performed work. 
In the stationary state the free energy is constant, and the 
dissipated energy and the performed work are equal.

Using Eqs. (\ref{19},\ref{19a}) we get
\begin{eqnarray}
\label{s1}
&&\dot{S}_{tot} =\gamma\, \frac{\kappa }{T_2\Gamma _1} \langle
 [ \delta F_2 ]^2 \rangle _1 -
\gamma^2\frac{\kappa }{\Gamma _1}\langle
 \delta F_2\,\partial _2 A\rangle_0 , \nonumber \\
&&\dot{ \Pi } = \gamma\,\frac{\kappa }{\Gamma _1} \langle
 [ \delta F_2 ]^2 \rangle _1-
\gamma^2\frac{\kappa (2T_2-T_1)}{\Gamma _1} \langle
 \delta F_2\,\partial _2 A \rangle_0 ,
\end{eqnarray}
where $\kappa =(T_1-T_2)/T_1$, and $\langle ... \rangle _{0(1)}$
 means averaging by the distribution $P_{0(1)}$.
We observe the following deceivingly simple relation,
valid to leading order in $\gamma$,
\begin{equation}
\label{s2}
\dot{ \Pi}
=T_2 \dot{ S}_{tot}+{\cal O}(\gamma^2)
\end{equation}
For a 
usual non-stationary system tending to equilibrium we have the 
following
relation between entropy production and energy dissipation:
$\dot{\Pi}=T\dot{S}_{tot}$, where $T$ is the temperature of the 
unique
thermal bath.  
On the other hand, Eq. (\ref{s2}) reflects degradation of the 
energy in the stationary state.
The distinguished role of $T_2$ is connected with conservation 
of detailed balance for small time-scales (see Eq.(\ref{3})).
Indeed, both entropy production and energy dissipation are small 
 on the characteristic times of the $x_1$-variable (\ref{s1}).
This equation also shows that when $T_2$ is close to zero,
 the energy dissipation (but not the entropy production)
 looses its leading term. At least in this limit the
$\gamma^2$-correction to $\dot{S}_{tot}$ is negative.

Let us apply the obtained general results to a simple toy model.
 We consider a pair of weakly-interacting oscillators with
 coordinates $x_1$, $x_2$
and Hamiltonian: $H=ax_1^2/2+ax_2^2/2+gx_1^2x_2^2$, where $a>0$, $g>0$. 
Very similar models are applied to describe an oscillator
with random frequency \cite{gardiner} or some electrical circuits 
\cite{strat}. 
For simplicity we shall discuss the model keeping only the first
non-vanishing order in the small parameter $g$.
The stationary distribution has the form (\ref{17}), with
\begin{equation}
\label{kk1}
A=\frac{g(T_1-T_2)}{a^2}(1-a\beta _2x_2^2)(1-a\beta _1x_1^2)
\end{equation}

After some calculations we get from (\ref{19}):
\begin{equation}
\label{25}
\dot{ S}_{tot}=\frac{8g^2}{\Gamma
_1}\frac{(T_1-T_2)^2}{a^3}(\gamma-\gamma^2), 
\end{equation}
\begin{equation}
\label{26}
\dot{ \Pi }=\frac{8g^2}{\Gamma _1}\frac{(T_1-T_2)^2}{a^3}(\gamma T_2+
\gamma^2 (T_1-2T_2))
\end{equation}
We see that in this model the $\gamma^2$ correction to
$\dot{S}_{tot}$ is always negative.
For $\dot{ \Pi }$ it is only the case if $T_1<2T_2$. 
We have thus provided a concrete example of our general results.

{\it (iv)} Let us now discuss the Onsager relations 
concerning heat transfer in the glassy state. 
These fundamental and experimentally testable relations 
were proposed by Onsager to describe transport in 
weakly non-equilibrium systems (the linear case)
\cite{strat}\cite{kampen}. 
Later they were generalized to the non-linear regime. 
Following standard arguments 
\cite{strat}\cite{kampen} the Onsager relation reads in our case 
\begin{equation}
\label{o1}
\partial _{\beta _1}\dot{Q}_2=\partial _{\beta _2}\dot{Q}_1,
\end{equation}
where  $\dot{Q}_i$, given by Eq. (\ref{urn1}), but see also Eq.
(\ref{f1}), 
is the heat flux from the thermal bath $i$. In the stationary case
one has  $\dot{Q}_1+\dot{Q}_2=0$. 
For our toy model we get from Eqs. (\ref{19},\ref{kk1}) 
\begin{equation}
\label{o5}
\dot{Q}_2=\gamma\,
\frac{g^2}{\Gamma _1a^3}\frac{\beta _1-\beta _2}{\beta _1^2 \beta _2^2}
+{\cal O}(\gamma^2). \end{equation}
The linear case corresponds to Eq. (\ref{o1}) with
$\beta _1\approx\beta _2$.
Indeed, then the fluxes can be written in more familiar form: 
$\dot{Q}_i=\sum_j L_{ij}\Delta \beta _j$, where 
$\Delta \beta _i=\beta _i-\beta _0$  is a small deviation
of the inverse temperature $\beta _i$ from its equilibrium value $\beta
_0$, and the $L_{ij}$ depend only on $\beta_0$ but not 
on $\beta _{1,2}$ separately. 
In that case the relation (\ref{o1}) takes the form: $L_{12}=L_{21}$
($=8g^2T_0^4/\Gamma_2a^3$ in our toy model).
Let us stress that this form of Onsager relations is applicable 
{\it only} for the linear case \cite{strat}\cite{kampen}, in 
contrast to the more general relation (\ref{o1}).

In fact the general validity of Eq. (\ref{o1}) in the linear regime
 is a fundamental theorem \cite{strat} supported by very general
 arguments. It means that 
{\it any breaking} of Eq. (\ref{o1}) can be connected only with 
$T_1\not =T_2$. The converse is not true: there are physically
 important  cases when
the Onsager relations hold for $T_1\not =T_2$
 \cite{strat}\cite{kampen}.
 Thus checking these relations for our {\it concrete} 
class of non-equilibrium systems seems very important. From 
Eqs. (\ref{19}), (\ref{19a}), (\ref{urn1}) we obtain to
leading order  in $\gamma$
\begin{eqnarray}
\label{o2}
&&\partial _{\beta _1}\dot{Q}_2-\partial _{\beta _2}\dot{Q}_1=\nonumber \\
&&\gamma\,\frac{\beta _1-\beta _2}{\Gamma _1}\,\{
\partial _{\beta _1}(T_2 \langle
 (\delta F_2)^2\rangle ) +\partial _{\beta _2}(T_2 \langle 
(\delta F_2)^2\rangle ) \}
\end{eqnarray}
In the linear regime with $\beta _1\approx\beta _2$ the Onsager relation
is satisfied automatically. 
However, for the considered glassy system it is the exceptional 
case, and (\ref{o1}) cannot be true in general. Indeed, 
for our toy model we get
\begin{equation}
\label{o4}
\partial _{\beta _1}\dot{Q}_2-\partial _{\beta _2}\dot{Q}_1=
\gamma\,\frac{16g^2}{\Gamma _1a^3}T_1T_2(T_1^2-T_2^2).
\end{equation}
implying a violation of the Onsager relation for any $T_1\not =T_2$,
because, due to (\ref{o5}), 
 the right hand side of (\ref{o4}) has the same order 
of magnitude as the individual terms in the left hand side. 
In this sense the violation is strong.
One can give also general, model-independent arguments supporting
breakdown of (\ref{o1}). Indeed, if it were to be  valid, Eq. (\ref{o2})
says that we should have
$\langle (\delta F_2)^2\rangle =\beta _2 f(\beta _1 -\beta _2)$ for 
all $\beta _1$, $\beta _2$, where $f$ is some  positive function.
Such a form cannot hold for a trivial reason: 
taking the limit $\beta _2\to 0$ one obtains zero in the 
right-hand side, while the left-hand side typically diverges,
or at least stays finite and non-zero including non-typical cases.
Our prediction for the breakdown of the Onsager relations 
in the glassy state should be testable experimentally. 
In glasses $T_2$ will correspond to effectively 
generated temperature \cite{theoehren}-\cite{theolongtermo}, and $T_1$
is the temperature of the environment. 
By changing the cooling rate,  also for these systems 
Eq. (\ref{o1}) constitutes a relation between measurable quantities.
A breaking of the Onsager relations can be investigated also
in this more realistic setup. The details will be reported elsewhere.

In conclusion, we have considered a stochastic model 
which contains all necessary ingredients of steady glassy behavior.
In spite of the fact that such a system can be very far 
from equilibrium, it allows a thermodynamic description. 
Generalizing the usual equilibrium thermodynamics,
we show that the glassy one can
be obtained by minimization of the energy, keeping all 
entropies fixed. Universal relations, Eqs. (\ref{s1}, \ref{s2}),
are obtained between entropy production and energy dissipation.
Energy dissipation (in contrast to entropy production)
is closely connected to the fluctuations of slow degrees of freedom, 
and it looses its leading term when the corresponding temperature 
is close to zero. We discuss cases where the corrections arising from a
finite but small ratio of characteristic times
lead to a decrease of
the entropy production and/or energy dissipation.
We show that the nonlinear Onsager relation for heat transfer in the
steady glassy state is always broken, reflecting its strongly 
non-equilibrium character. As the effect is of order unity, 
this breaking should be testable experimentally. 
It is also reminiscent of
the breakdown of the Maxwell and Ehrenfest relations in the glassy 
state~\cite{theojpa}\cite{theoehren}.

A.E. A. is grateful to FOM (The Netherlands) for financial support.

\end{multicols}
\end{document}